\documentclass[a4paper]{article} 

\usepackage{amsmath, amssymb} 
\usepackage{mathrsfs} 
\usepackage{graphics,epsfig}
\usepackage{qip}

\usepackage{underscore}
\usepackage{authblk}

\usepackage{hyperref}  
\usepackage{env4article}
\usepackage{eprint}

\title{On the Power of Quantum Encryption Keys}

\author{Akinori Kawachi\email{kawachi@is.titech.ac.jp}\ }
\author{Christopher Portmann\email{portmann.c.aa@m.titech.ac.jp}}

\affil{\small Department of Mathematical and Computing Sciences, Tokyo Institute of Technology, 2-12-1 Ookayama, Meguro-ku, Tokyo 152-8552, Japan.}

\date{\small August 21, 2008}

\begin{document}

\maketitle

\begin{abstract}
  The standard definition of quantum state randomization, which is the
  quantum analog of the classical one-time pad, consists in applying
  some transformation to the quantum message conditioned on a
  classical secret key $k$. We investigate encryption schemes in which
  this transformation is conditioned on a quantum encryption key state
  $\rho_k$ instead of a classical string, and extend this
  symmetric-key scheme to an asymmetric-key model in which copies of
  the same encryption key $\rho_k$ may be held by several different
  people, but maintaining information-theoretical security.

  We find bounds on the message size and the number of copies of the
  encryption key which can be safely created in these two models in
  terms of the entropy of the decryption key, and show that the
  optimal bound can be asymptotically reached by a scheme using
  classical encryption keys.

  This means that the use of quantum states as encryption keys does
  not allow more of these to be created and shared, nor encrypt larger
  messages, than if these keys are purely classical.
\end{abstract}


\section{Introduction}
\label{sec:intro}

\subsection{Quantum Encryption}
\label{sec:intro.qenc}

To encrypt a quantum state $\sigma$, the standard procedure consists
in applying some (unitary) transformation $U_k$ to the state, which
depends on a classical string $k$. This string serves as secret key,
and anyone who knows this key can perform the reverse operation and
obtain the original state. If the transformations $U_1, U_2, \dotsc$
are chosen with probabilities $p_1, p_2, \dotsc$, such that when
averaged over all possible choices of key,
\begin{equation}
  \label{eq:intro.qsr}
  \mathcal{R}(\sigma) = \sum_k p_k U_k \sigma \hconj{U_k},
\end{equation}
the result looks random, i.e., close to the fully mixed state,
$\mathcal{R}(\sigma) \approx \id/d$, this cipher can safely be
transmitted on an insecure channel. This procedure is called
\emph{approximate quantum state randomization} or \emph{approximate
  quantum one-time pad}~\cite{HLSW04,AS04,DN06} or \emph{quantum
  one-time pad}, \emph{quantum Vernam cipher} or \emph{quantum private
  channel} in the case of perfect security~\cite{BR03,AMTW00,NS07},
and is the quantum equivalent of the classical one-time pad.

An encryption scheme which uses such a randomization procedure is
called \emph{symmetric}, because the same key is used to encrypt and
decrypt the message. An alternative paradigm is \emph{asymmetric-key
  cryptography}, in which a different key is used for encryption and
decryption. In such a cryptosystem the encryption key may be shared
amongst many different people, because possessing this key is not
sufficient to perform the reverse operation, decryption. This can be
seen as a natural extension of symmetric-key cryptography, because
this latter corresponds to the special case in which the encryption
and decryption keys are identical and can be shared with only one
person.

Although the encryption model given in \eqnref{eq:intro.qsr} is
symmetric, by replacing the classical encryption key with a quantum
state we can make it asymmetric. To see this, let us rewrite
\eqnref{eq:intro.qsr} as
\begin{equation}
  \label{eq:intro.qsr2}
  \mathcal{R}(\sigma) = \sum_k p_k \tr_K\left[ U \left(\proj{k}^K \tensor \sigma^S \right)\hconj{U} \right],
\end{equation}
where $U \coloneqq \sum_k \proj{k} \tensor U_k$. The encryption key in
\eqnref{eq:intro.qsr2}, $\proj{k}$, is diagonal in the computational
basis, i.e., classical, but an arbitrary quantum state, $\rho_k$,
could be used instead, e.g.,
\begin{equation}
  \label{eq:intro.qsr3}
  \mathcal{R}(\sigma) = \sum_k p_k \tr_K\left[ U \left(\rho_k^K \tensor \sigma^S \right)\hconj{U} \right],
\end{equation}
for some set of quantum encryption keys $\{\rho_k\}_k$.

If the sender only holds such a quantum encryption key state $\rho_k$
without knowing the corresponding decryption key $k$, then the
resulting model is asymmetric in the sense that possessing this copy
of the encryption key state is enough to perform the encryption, but
not to decrypt. So many different people can hold copies of the
encryption key without compromising the security of the scheme. It is
generally impossible to distinguish between non-orthogonal quantum
states with certainty (we refer to the textbook by Nielsen and
Chuang~\cite{NC00} for an introduction to quantum information), so
measuring a quantum state cannot tell us precisely what it is, and
possessing a copy of the encryption key state does not allow us to
know how the quantum message got transformed, making it impossible to
guess the message, except with exponentially small probability.

Up to roughly $\log N$ copies of a state can be needed to discriminate
between $N$ possible states~\cite{HW06}, so such a scheme could allow
the same encryption key to be used several times, if multiple copies
of this quantum key state are shared with any party wishing to encrypt
a message. The scheme will stay secure as long as the number of copies
created stays below a certain threshold. What is more, the security
which can be achieved is information-theoretic like for standard
quantum state randomization schemes~\cite{HKK08}, not computational
like most asymmetric-key encryption schemes.

Such an asymmetric-key cryptosystem is just a possible application of
a quantum state randomization scheme which uses quantum keys. It is
also interesting to study quantum state randomization with quantum
keys for itself (in the symmetric-key model), without considering
other parties holding extra copies of the same encryption key. In this
paper we study these schemes in both the symmetric-key and asymmetric-key
models, and compare their efficiency in terms of message size and
number of usages of the same encryption key to quantum state
randomization schemes which use only classical keys.

\subsection{Related Work}

Quantum one-time pads were first proposed in \cite{BR03,AMTW00} for
perfect security, then approximate security was considered in, e.g., 
\cite{HLSW04,AS04,DN06}. All these schemes assume the sender and
receiver share some secret classical string which is used only once to
perform the encryption. We extend these models in the symmetric-key
case by conditioning the encryption operation on a quantum key and
considering security with multiple uses of the same key, and then in
the asymmetric-key case by considering security with multiple users
holding copies of the same encryption key.

The first scheme using quantum keys in an asymmetric-key model was
proposed by Kawachi et al.~\cite{KKNY05}, although they considered the
restricted scenario of classical messages. Their scheme can encrypt a
$1$ bit classical message, and their security proof is computational,
as it reduces the task of breaking the scheme to a graph automorphism
problem. They extended their scheme to a multi-bit
version~\cite{KKNY06}, but without security proof. Hayashi et
al.~\cite{HKK08} then gave an information-theoretical security proof
for \cite{KKNY06}. The quantum asymmetric-key model we consider is a
generalization and extension of that of~\cite{KKNY05,KKNY06}.

\subsection{Main Contributions}

The main result of this paper is that using quantum encryption keys
has no advantage over classical keys with respect to the number of
copies of the encryption key which can be safely created and to the
size of the messages which can be encrypted, both in the symmetric and
asymmetric-key models. Contrary to what was believed and motivated
previous works with quantum keys, the intrinsic indistinguishability
of quantum states does not allow more of these to be created and
shared as encryption keys, than if these keys are purely
classical.

To show this, we first find an upper bound on the quantum message size
and on the number of copies of the encryption key which can be
securely produced. We show that if $t$ copies of the key are created
and if the quantum messages encrypted are of dimension $d$, then they
have to be such that $t \log d \lesssim \Hh \left( \mathcal{K} \right)$
for the scheme to be secure, where $\Hh \left( \mathcal{K} \right)$ is
the entropy of the decryption key.

We then construct a quantum state randomization scheme and show that
it meets this upper bound in both the symmetric and asymmetric-key
models. The encryption keys this scheme uses are however all diagonal
in the same bases, i.e., classical. This means that the scheme with
classical keys is optimal in terms of message size and number of
usages of the same key, and no scheme with quantum keys can perform
better.

We also show how to extend quantum asymmetric-key encryption schemes
for classical message (such as \cite{KKNY06}) to encrypt quantum
messages as well. To do this, we combine these schemes for classical
messages with a standard quantum one-time pad, and prove that the
resulting scheme is still secure.

\subsection{Organization of the Paper}

In \secref{sec:model} we develop the encryption models with quantum
keys sketched in this introduction. We first redefine quantum state
randomization schemes using quantum keys instead of classical keys in
\secref{sec:model.quantum} and generalize the standard security
definition for multiple usage of the same key in this symmetric-key
model. In \secref{sec:model.cryptosystem} we then show how to
construct an asymmetric-key cryptosystem using such a quantum state
randomization scheme with quantum keys and define its
security. \secref{sec:model.classical} contains a few notes about the
special case of classical messages, which are relevant for the rest of
the paper.

In \secref{sec:keysize} we find an upper bound on the message size
and number of copies of the encryption key which can be created, both
for the symmetric and asymmetric-key models.

In \secref{sec:scheme} we construct a quantum state
randomization scheme which uses classical encryption keys, but which
meets the optimality bounds for quantum keys from the previous section
in both models. We give this construction in three steps. First in
\secref{sec:scheme.classical} we construct a scheme which can
randomize classical messages only. Then in
\secref{sec:scheme.quantum} we show how to combine this scheme for
classical messages with a standard approximate quantum one-time pad to
randomize any quantum state. And finally in \secref{sec:scheme.key}
we calculate the key size of the scheme proposed and show that it
corresponds to the bound found in \secref{sec:keysize}.

We conclude in \secref{sec:consequence} with a brief summary and
further comments about the results.

\section{Encryption Model}
\label{sec:model}

\subsection{Quantum Encryption Keys}
\label{sec:model.quantum}

Let us consider a setting in which we have two parties, a sender and a
receiver, who wish to transmit a quantum state, $\sigma$, from one to
the other in a secure way over an insecure channel. If they share a
secret classical string, $k$, they can apply some completely positive,
trace-preserving (CPTP) map $\mathcal{E}_k$ to the quantum message and
send the cipher $\mathcal{E}_k(\sigma)$. If the key $k$ was chosen
with probability $p_k$, to any person who does not know this key the
transmitted state is
\begin{equation}
  \label{eq:model.quantum.qsr}
  \mathcal{R}(\sigma) = \sum_k p_k \mathcal{E}_k(\sigma),
\end{equation}
which will look random for ``well chosen'' maps $\mathcal{E}_k$. This
is the most general from of quantum state randomization~\cite{NS07}.

If instead the sender has a quantum state $\rho_k$, he can apply some
CPTP map $\mathcal{E}$ to both the shared state and the quantum
message, and send $\mathcal{E}(\rho_k \tensor \sigma)$. So for someone
who does not know $\rho_k$ the state sent is
\begin{equation}
  \label{eq:model.quantum.qsr2}
  \mathcal{R}(\sigma) = \sum_k p_k \mathcal{E}(\rho_k \tensor \sigma).
\end{equation}

It is clear that \eqnsref{eq:model.quantum.qsr} and
\eqref{eq:model.quantum.qsr2} produce equivalent ciphers, because for
every set of CPTP maps $\{\mathcal{E}_k\}_k$ there exists a map
$\mathcal{E}$ and set of states $\{\rho_k\}_k$ such that for all
messages $\sigma$, $\mathcal{E}_k(\sigma) = \mathcal{E}(\rho_k \tensor
\sigma)$, and vice versa. The difference lies in the knowledge needed
to perform the encryption. In the first case
(\eqnref{eq:model.quantum.qsr}) the sender needs to know the secret
key $k$ to know which CPTP map $\mathcal{E}_k$ to apply. In the second
case (\eqnref{eq:model.quantum.qsr2}) the sender only needs to hold a
copy of the encryption key $\rho_k$, he does not need to know what it
is or what secret key $k$ it corresponds to. This allows us to
construct in \secref{sec:model.cryptosystem} a quantum asymmetric-key
cryptosystem in which copies of the same encryption key $\rho_k$ can
be used by many different users. In this section we focus on the
symmetric-key model and define quantum state randomization (QSR)
schemes with quantum encryption keys and their security in this model.

\begin{deff}
  \label{def:model.quantum.enc}
  Let $\mathcal{B}(\hilbert)$ denote the set of linear operators on
  \hilbert.

  A quantum state randomization (QSR) scheme with quantum
  encryption keys consists of the following tuple, \[\mathbb{T} =
  \left(P_\mathcal{K}, \{\rho_k\}_{k \in \mathcal{K}}, \mathcal{E}
  \right).\]

  $\rho_k \in \mathcal{B}(\hilbert_K)$ are density operators on a
  Hilbert space $\hilbert_K$. They are called \emph{encryption keys}
  and are indexed by elements $k \in \mathcal{K}$ called
  \emph{decryption keys}.

  $P_\mathcal{K}(\cdot)$ is a probability distribution over the set of
  decryption keys $\mathcal{K}$, corresponding to the probability with
  which each en/decryption key-pair should be chosen.

  $\mathcal{E} : \mathcal{B}(\hilbert_K \tensor
  \hilbert_S) \rightarrow \mathcal{B}(\hilbert_C)$, is a completely
  positive, trace-preserving (CPTP) map from the set of linear
  operators on the joint system of encryption key and message Hilbert
  spaces, $\hilbert_K$ and $\hilbert_S$ respectively, to the set of
  linear operators on the cipher Hilbert space $\hilbert_C$, and is
  called \emph{encryption operator}.

  To encrypt a quantum message given by its density operator $\sigma
  \in \mathcal{B}(\hilbert_S)$ with the encryption key $\rho_k$, the
  encryption operator is applied to the key and message, resulting in
  the cipher \[\rho_{k,\sigma} \coloneqq \mathcal{E}(\rho_k \tensor
  \sigma).\]
\end{deff}

\defref{def:model.quantum.enc} describes how to encrypt a quantum message,
but for such a scheme to be useful, it must also be possible to
decrypt the message for someone who knows which key $k$ was used,
i.e., it must be possible to invert the encryption operation.

\begin{deff}
  \label{def:model.quantum.dec}
  A QSR scheme given by the tuple $\mathbb{T} = \left(P_\mathcal{K},
    \{\rho_k\}_{k \in \mathcal{K}}, \mathcal{E} \right)$ is said to be
  \emph{invertible on the set $\mathcal{S} \subseteq
    \mathcal{B}(\hilbert_S)$} if for every $k \in \mathcal{K}$ with
  $P_{\mathcal{K}}(k) > 0$ there exists a CPTP map $\mathcal{D}_k :
  \mathcal{B}(\hilbert_C) \rightarrow \mathcal{B}(\hilbert_S)$ such
  that for all density operators $\sigma \in
  \mathcal{S}$, \[ \mathcal{D}_k \mathcal{E} (\rho_k
  \tensor \sigma) = \sigma. \]
\end{deff}

Furthermore, a QSR scheme must -- as its name says -- randomize a
quantum state. We define this in the same way as previous works on
approximate quantum state randomization~\cite{HLSW04,AS04,DN06}, by
bounding the distance between the ciphers averaged over all possible
choices of key and some state independent from the message. We however
generalize this to encrypt $t$ messages with the same key, because the
asymmetric-key model we define \secref{sec:model.cryptosystem} will
need this. It is always possible to consider the case $t = 1$ in the
symmetric-key model, if multiple uses of the same key are not desired.

We will use the trace norm as distance measure between two states,
because it is directly related to the probability that an optimal
measurement can distinguish between these two states, and is therefore
meaningful in the context of eavesdropping. The trace norm of a matrix
$A$ is defined by $\trnorm{A} \coloneqq \tr{|A|} =
\tr{\sqrt{\hconj{A}A}}$, which is also equal to the sum of the
singular values of $A$.

\begin{deff}
  \label{def:model.quantum.security}
  A QSR scheme given by the tuple $\mathbb{T} = \left(P_\mathcal{K},
    \{\rho_k\}_{k \in \mathcal{K}}, \mathcal{E} \right)$ is said to be
  \emph{$(t,\epsilon)$-randomizing on the set $\mathcal{S} \subseteq
    \mathcal{B}(\hilbert_S)$} if there exists a density operator $\tau
  \in \mathcal{B}\left(\hilbert_C^{\tensor t}\right)$ such that for
  all $t$-tuples of message density operators $\omega =
  (\sigma_1,\dotsc,\sigma_t) \in \mathcal{S}^{\times t}$
  \begin{equation}
    \label{eq:model.quantum.security}
    \trnorm{ \mathcal{R}(\omega) - \tau} \leq \epsilon,
  \end{equation}
  where $\mathcal{R}(\omega) = \sum_k P_\mathcal{K}(k)
  \rho_{k,\sigma_1} \tensor \dotsb \tensor \rho_{k,\sigma_t}$ and
  $\rho_{k,\sigma_i} = \mathcal{E}(\rho_k \tensor \sigma_i)$.
\end{deff}

\subsection{Quantum Asymmetric-Key Cryptosystem}
\label{sec:model.cryptosystem}

\begin{figure}[htb]
  \begin{center}
    \resizebox{11cm}{!}{
      \includegraphics*{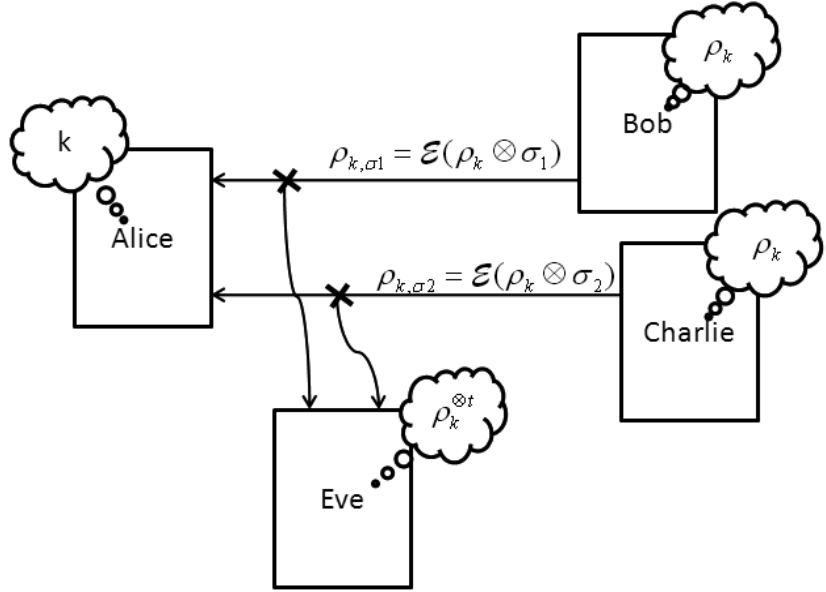} 
    }
  \end{center}
  \caption[Quantum asymmetric-key cryptosystem model]{\emph{Quantum
      asymmetric-key cryptosystem model.} Bob and Charlie hold copies
    of Alice's encryption key $\rho_k$. To send her a message, they
    encrypt it with the key and a given QSR scheme, and send the
    resulting cipher to her. An eavesdropper, Eve, may intercept the
    ciphers as well as possess some copies of the encryption key
    herself.}
  \label{fig:model}
\end{figure}

As announced in the previous section, the idea behind the quantum
asymmetric-key cryptosystem model is that many different people hold a
copy of some quantum state $\rho_k$ which serves as encryption key,
and anyone who wishes to send a message to the originator of the
encryption keys uses a quantum state randomization scheme, as
described in \defref{def:model.quantum.enc}. This is depicted in
\figref{fig:model}.

If the QSR scheme used to encrypt the messages is
$(t,\epsilon)$-randomizing and no more than $t$ copies of the
encryption key were released, an eavesdropper who intercepts the
ciphers will not be able to distinguish them from some state
independent from the messages, so not get any information about these
messages. This is however not the only attack he may perform.

As we consider a scenario in which copies of the encryption key are
shared between many different people, the adversary could hold one or
many of them. If a total of $t$ copies of the encryption key were
produced and $t_1$ were used to encrypt messages $\omega =
(\sigma_1,\dotsc,\sigma_{t_1})$, in the worst case we have to assume
that the adversary has the $t_2 \coloneqq t - t_1$ remaining unused
copies of the key. So his total state
is \begin{equation} \label{eq:model.cryptosystem.adversary}
  \rho^E_\omega \coloneqq \sum_{k \in \mathcal{K}} P_{\mathcal{K}}(k)
  \rho_{k,\sigma_1} \tensor \dotsb \tensor \rho_{k,\sigma_{t_1}}
  \tensor \rho_k^{\tensor t_2},\end{equation} where
$\rho_{k,\sigma_i}$ is the cipher of the message $\sigma_i$ encrypted
with the key $\rho_k$. This leads to the following security
definition.

\begin{deff}
  \label{def:model.cryptosystem}
  We call a quantum asymmetric-key cryptosystem
  \emph{$(t,\epsilon)$-in\-dis\-tin\-guish\-able on the set
    $\mathcal{S} \subseteq \mathcal{B}(\hilbert_S)$} if for all $t_1
  \in \{0,1,\dotsc,t\}$, $t_2 \coloneqq t - t_1$, there exists a
  density operator $\tau \in \mathcal{B}\left(\hilbert_C^{\tensor t_1}
    \tensor \hilbert_K^{\tensor t_2}\right)$ such that for all
  $t_1$-tuples of message density operators $\omega =
  (\sigma_1,\dotsc,\sigma_{t_1}) \in \mathcal{S}^{\times
    t_1}$, \[\trnorm{\rho^E_\omega - \tau} \leq \epsilon,\] where
  $\rho^E_\omega$ is the state the adversary obtains as defined in
  \eqnref{eq:model.cryptosystem.adversary}.
\end{deff}

\begin{rem}
  \label{rem:model.cryptosystem}
  \defref{def:model.cryptosystem} is clearly more general than the
  security criteria of \defref{def:model.quantum.security}
  ($(t,\epsilon)$-randomization) as this latter corresponds to the
  special case $t_1 = t$. However, for the scheme constructed in
  \secref{sec:scheme} the two are equivalent, and proving one proves
  the other. This is the case in particular if the encryption key is
  equal to the cipher of some specific message $\sigma_0$, i.e.,
  $\rho_k = \rho_{k,\sigma_0} = \mathcal{E}(\rho_k \tensor \sigma_0)$,
  in which case holding an extra copy of the encryption key does not
  give more information about the decryption key than holding an extra
  cipher state.
\end{rem}

\subsection{Classical Messages}
\label{sec:model.classical}

In the following sections we will also be interested in the special
case of schemes which encrypt classical messages only. Classical
messages can be represented by a set of mutually orthogonal quantum
states, which we will take to be the basis states of the message
Hilbert space and denote by $\{\ket{s}\}_{s \in \mathcal{S}}$. So these schemes
must be invertible and randomizing on the set of basis states of the
message Hilbert space.

When considering classical messages only, we will simplify the
notation when possible and represent a message by a string $s$ instead
of by its density matrix $\proj{s}$, e.g., the cipher of the message
$s$ encrypted with the key $\rho_k$ is \[\rho_{k,s} \coloneqq
\mathcal{E} \left( \rho_k \tensor \proj{s} \right).\]

\begin{rem}
  \label{rem:model.classical.dec}
  \defref{def:model.quantum.dec} (invertibility) can be simplified
  when only classical messages are considered: a QSR scheme given by
  the tuple $\mathbb{T} = \left(P_\mathcal{K}, \{\rho_k\}_{k \in
      \mathcal{K}}, \mathcal{E} \right)$ is invertible for the set of
  classical messages $\mathcal{S}$, if for every $k \in \mathcal{K}$ with
  $P_{\mathcal{K}}(k) > 0$ the ciphers $\{\rho_{k,s}\}_{s \in \mathcal{S}}$ are
  mutually orthogonal, where $\rho_{k,s} \coloneqq \mathcal{E} \left(
    \rho_k \tensor \proj{s} \right)$ for some orthonormal basis
  $\{\ket{s}\}_{s \in \mathcal{S}}$ of the message Hilbert space $\hilbert_S$.
\end{rem}

We will also use a different but equivalent definition to measure how
well a scheme can randomize a message when dealing with classical
messages. This new security criteria allows us to simplify some proofs.

\begin{deff}
  \label{def:model.classical.security}
  A QSR scheme given by the tuple $\mathbb{T} = \left(P_\mathcal{K},
    \{\rho_k\}_{k \in \mathcal{K}}, \mathcal{E} \right)$ is said to be
  \emph{$(t,\epsilon)$-secure for the set of classical messages $\mathcal{S}$}
  if for all probability distributions $P_{\mathcal{S}^t}(\cdot)$ over the set
  of $t$-tuples of messages $\mathcal{S}^{\times t}$,
  \begin{equation}
    \label{eq:model.classical.security}
    \trnorm{\rho^{S^tC^t} - \rho^{S^t} \tensor \rho^{C^t}} \leq \epsilon,
  \end{equation}
  where $\rho^{S^tC^t}$ is the state of the joint systems of $t$-fold
  message and cipher Hilbert spaces, and $\rho^{S^t}$ and $\rho^{C^t}$
  are the result of tracing out the cipher respectively message
  systems. I.e.,
  \begin{align*}
    \rho^{S^tC^t} & = \sum_{s \in \mathcal{S}^{\times t}}
    P_{\mathcal{S}^t}(s) \proj{s} \tensor \sum_{k \in \mathcal{K}}
    P_{\mathcal{K}}(k) \rho_{k,s_1} \tensor \dotsb
    \tensor \rho_{k,s_t}, \\
    \rho^{S^t} & = \sum_{s \in
      \mathcal{S}^{\times t}} P_{\mathcal{S}^t}(s) \proj{s}, \\
    \rho^{C^t} & = \sum_{s \in \mathcal{S}^{\times t}}
    P_{\mathcal{S}^t}(s) \sum_{k \in \mathcal{K}} P_{\mathcal{K}}(k)
    \rho_{k,s_1} \tensor \dotsb \tensor \rho_{k,s_t},
  \end{align*}
  where $s = (s_1,\dotsc,s_t)$.
\end{deff}

This security definition can be interpreted the following way. No
matter what the probability distribution on the secret messages is --
let the adversary choose it -- the message and cipher spaces are
nearly in product form, i.e., the cipher gives next to no information
about the message.

The following lemma proves that this new security definition is
equivalent to the previous one (\defref{def:model.quantum.security})
up to a constant factor.

\begin{lem}
  \label{lem:model.classical.security}
  If a QSR scheme is $(t,\epsilon)$-randomizing for a set of classical
  messages $\mathcal{S}$, then it is $(t,2\epsilon)$-secure for
  $\mathcal{S}$. If a QSR scheme is $(t,\epsilon)$-secure for a set of
  classical messages $\mathcal{S}$, then it is
  $(t,2\epsilon)$-randomizing for $\mathcal{S}$.
\end{lem}

\begin{proof}
  In order to simplify the notation we will set $s \coloneqq
  (s_1,\dotsc,s_t)$ and $\rho_{k,s} \coloneqq \rho_{k,s_1} \tensor
  \dotsb \tensor \rho_{k,s_t}$. The left-hand side of
  \eqnref{eq:model.classical.security} can then be rewritten as
  \begin{align}
    & \trnorm{\rho^{S^tC^t} - \rho^{S^t} \tensor \rho^{C^t}} \notag \\ & \qquad = \left\| \sum_{s \in \mathcal{S}^{\times t}} P_{\mathcal{S}^t}(s) \proj{s} \tensor \sum_{k \in \mathcal{K}} P_{\mathcal{K}}(k) \rho_{k,s} \right. \notag \\ & \qquad \qquad - \left. \sum_{s \in \mathcal{S}^{\times t}} P_{\mathcal{S}^t}(s) \proj{s} \tensor \sum_{\substack{r \in \mathcal{S}^{\times t} \\ k \in \mathcal{K}}} P_{\mathcal{K}}(k) P_{\mathcal{S}^t}(r) \rho_{k,r} \right\|_\text{tr} \notag \\
    & \qquad = \sum_{s \in \mathcal{S}^{\times t}} P_{\mathcal{S}^t}(s) \trnorm{ \sum_{k \in \mathcal{K}} P_{\mathcal{K}}(k) \rho_{k,s} - \sum_{\substack{r \in \mathcal{S}^{\times t} \\ k \in \mathcal{K}}} P_{\mathcal{K}}(k) P_{\mathcal{S}^t}(r) \rho_{k,r}}. \label{eq:model.classical.security.lem.1}
  \end{align}
  If this must be less than $\epsilon$ for all probability
  distributions $P_{\mathcal{S}^t}$ then for the distribution
  $P_{\mathcal{S}^t}(s_1) = P_{\mathcal{S}^t}(s_2) = 1/2$ for any two
  elements $s_1,s_2 \in \mathcal{S}^{\times t}$ we have from
  \eqnref{eq:model.classical.security.lem.1}
  \[\frac{1}{2} \trnorm{ \sum_{k \in \mathcal{K}} P_{\mathcal{K}}(k) \rho_{k,s_1} -  \sum_{k \in \mathcal{K}} P_{\mathcal{K}}(k) \rho_{k,s_2}} \leq \epsilon. \]
  This immediately implies $(t,2\epsilon)$-randomization.

  To prove the converse we apply the triangle inequality to
  \eqnref{eq:model.classical.security.lem.1} and get
  \begin{multline*}\trnorm{\rho^{S^tC^t} - \rho^{S^t} \tensor \rho^{C^t}} \leq \\
  \sum_{s \in \mathcal{S}^{\times t}} P_{\mathcal{S}^t}(s) \sum_{r \in
    \mathcal{S}^{\times t}} P_{\mathcal{S}^t}(r) \trnorm{ \sum_{k \in
      \mathcal{K}} P_{\mathcal{K}}(k) \rho_{k,s} - \sum_{k \in
      \mathcal{K}} P_{\mathcal{K}}(k) \rho_{k,r}}.\end{multline*} By the definition
  of $(t,\epsilon)$-randomization
  (\defref{def:model.quantum.security}) and the triangle inequality we
  know that \[\trnorm{ \sum_{k \in \mathcal{K}} P_{\mathcal{K}}(k)
    \rho_{k,s} - \sum_{k \in \mathcal{K}} P_{\mathcal{K}}(k)
    \rho_{k,r}} \leq 2 \epsilon,\] for all $r,s \in \mathcal{S}^{\times t}$, which concludes the proof. 
\end{proof}

\section{Lower bounds on the Key Size}
\label{sec:keysize}

It is intuitively clear that the more copies of the encryption key
state $\rho_k$ are created, the more information the adversary gets
about the decryption key $k \in \mathcal{K}$ and the more insecure the
scheme becomes. As it turns out, the number of copies of the
encryption key which can be safely used is directly linked to the size
of the decryption key, i.e., the cardinality of the decryption key set
$\mathcal{K}$.

Let us assume a QSR scheme with quantum encryption keys is used to
encrypt classical messages of size $m$. Then if $t$ copies of the
encryption key state are released and used, the size of the total
message encrypted with the same decryption key $k$ is $tm$. We prove
in this section that the decryption key has to be of the same size as
the total message to achieve information-theoretical security, i.e.,
$\log |\mathcal{K}| \gtrsim tm$. In \secref{sec:scheme} we then give
a scheme which reaches this bound asymptotically.

\begin{thm}
  \label{thm:keysize}
  If a QSR scheme given by the tuple $\mathbb{T} =
  \left(P_\mathcal{K}, \{\rho_k\}_{k \in \mathcal{K}}, \mathcal{E}
  \right)$ is invertible for the set of classical messages $\mathcal{S}$, then
  when $t$ messages $(s_1,\dotsc,s_t)$ are chosen from $\mathcal{S}$ with
  (joint) probability distribution $P_{\mathcal{S}^t}(s_1,\dotsc,s_t)$ and
  encrypted with the same key,
  \begin{equation} \label{eq:keysize.thm} \trnorm{\rho^{S^tC^t} -
      \rho^{S^t} \tensor \rho^{C^t}} \geq \frac{\Hh\left(\mathcal{S}^t\right) -
      \Hh\left(\mathcal{K}\right) - 2} {4 t \log |S|} , \end{equation} where
  $\Hh(\cdot)$ is the Shannon entropy and $\rho^{S^tC^t}$ is the state
  of the $t$-fold message and cipher systems:
  \begin{equation} \label{eq:keysize.thm.sys} \begin{split}
       \rho^{S^tC^t} & = \sum_{s \in \mathcal{S}^{\times t}}
    P_{\mathcal{S}^t}(s) \proj{s} \tensor \sum_{k \in \mathcal{K}}
    P_{\mathcal{K}}(k) \rho_{k,s_1} \tensor \dotsb
    \tensor \rho_{k,s_t}, \\
    \rho^{S^t} & = \sum_{s \in
      \mathcal{S}^{\times t}} P_{\mathcal{S}^t}(s) \proj{s}, \\
    \rho^{C^t} & = \sum_{s \in \mathcal{S}^{\times t}}
    P_{\mathcal{S}^t}(s) \sum_{k \in \mathcal{K}} P_{\mathcal{K}}(k)
    \rho_{k,s_1} \tensor \dotsb \tensor \rho_{k,s_t},
  \end{split} \end{equation}  where $s = (s_1,\dotsc,s_t)$.
\end{thm}

\begin{proof}
  A theorem by Alicki and Fanes~\cite{AF04} tells us that for any
  two states $\rho^{AB}$ and $\sigma^{AB}$ on the joint system
  $\hilbert_{AB} = \hilbert_A \tensor \hilbert_B$ with $\delta
  \coloneqq \trnorm{\rho^{AB} - \sigma^{AB}} \leq 1$ and $d_A \coloneqq \dim
  \hilbert_A$,
  \begin{equation}
    \label{eq:keysize.thm.1}
    \left| \Ss\left(\rho^{AB} \middle| \rho^B\right) -
      \Ss\left(\sigma^{AB}\middle|\sigma^B\right) \right| \leq 4 \delta \log d_A + 2 h \left(
      \delta \right),
  \end{equation}
  where $\Ss\left(\rho^{AB} \middle| \rho^B\right) \coloneqq
  \Ss\left(\rho^{AB}\right) - \Ss\left(\rho^B\right)$ is the conditional
  Von Neumann entropy and $h(p) \coloneqq p \log \frac{1}{p} + (1-p)
  \log \frac{1}{1-p}$ is the binary entropy. $h(\delta) \leq 1$,
  so from \eqnref{eq:keysize.thm.1} we get
  \begin{equation*}
    \label{eq:keysize.thm.2}
    \trnorm{\rho^{AB} - \sigma^{AB}} \geq \frac{ \left| \Ss\left(\rho^{AB} \middle| \rho^B\right) -
        \Ss\left(\sigma^{AB}\middle|\sigma^B\right) \right| - 2} {4 \log d_A}.
  \end{equation*}
  By applying this to the left-hand side of \eqnref{eq:keysize.thm} we obtain   \begin{equation*} 
    \label{eq:keysize.thm.3}
    \trnorm{\rho^{S^tC^t} - \rho^{S^t} \tensor \rho^{C^t}} \geq \frac{\Ss\left(\rho^{S^t}\right) + \Ss\left(\rho^{C^t}\right) - \Ss\left(\rho^{S^tC^t}\right)  - 2} {4 t \log |\mathcal{S}|}.
  \end{equation*}

  To prove this theorem it remains to show that \[\Ss\left(\rho^{S^t}\right) + \Ss\left(\rho^{C^t}\right) - \Ss\left(\rho^{S^tC^t}\right) \geq \Hh\left(\mathcal{S}^t\right) - \Hh\left(\mathcal{K}\right).\] For this we will need the two following bounds on the Von Neumann entropy (see e.g, \cite{NC00}):
  \begin{align*}
    \Ss \left(\sum_{x \in \mathcal{X}} p_x \rho_x \right) & \geq \sum_{x \in \mathcal{X}} p_x \Ss \left(\rho_x\right), \\
    \Ss \left(\sum_{x \in \mathcal{X}} p_x \rho_x \right) & \leq \Hh\left(\mathcal{X}\right) + \sum_{x \in \mathcal{X}} p_x \Ss \left(\rho_x\right).
  \end{align*}
  Equality is obtained in the second equation if the states
  $\{\rho_x\}_{x \in \mathcal{X}}$ are all mutually orthogonal. By using these bounds and \eqnref{eq:keysize.thm.sys} we see that
  \begin{align*}
    \Ss\left(\rho^{S^tC^t}\right) & = \Hh \left(\mathcal{S}^t\right) +
    \sum_{s \in \mathcal{S}^{\times t}}
    P_{\mathcal{S}^t}(s) \Ss \left( \sum_{k \in
        \mathcal{K}} P_{\mathcal{K}}(k) \rho_{k,s_1} \tensor \dotsb
      \tensor \rho_{k,s_t} \right) \\
    & \leq \Hh \left(\mathcal{S}^t\right) + \Hh
    \left(\mathcal{K}\right) + \sum_{\substack{s \in \mathcal{S}^{\times t} \\ k \in \mathcal{K}}}
    P_{\mathcal{K}}(k) P_{\mathcal{S}^t}(s)
     \Ss \left(\rho_{k,s_1} \tensor \dotsb \tensor
      \rho_{k,s_t}\right), \\
    \Ss\left(\rho^{S^t}\right)  & = \Hh \left( \mathcal{S}^t\right),\\
    \Ss\left(\rho^{C^t}\right) & \geq \sum_{k \in \mathcal{K}}
    P_{\mathcal{K}}(k) \Ss \left( \sum_{s \in \mathcal{S}^{\times t}} P_{\mathcal{S}^t}(s)\rho_{k,s_1}
      \tensor \dotsb \tensor
      \rho_{k,s_t}\right) \\
    & = \Hh \left( \mathcal{S}^t\right) + \sum_{\substack{s \in \mathcal{S}^{\times t} \\ k \in \mathcal{K}}}
    P_{\mathcal{K}}(k) P_{\mathcal{S}^t}(s)
     \Ss \left(\rho_{k,s_1} \tensor \dotsb \tensor
      \rho_{k,s_t}\right).
  \end{align*}
  We have equality in the last line because the scheme is invertible
  on $\mathcal{S}$, i.e., by \defref{def:model.quantum.dec} and
  \remref{rem:model.classical.dec} the states $\{\rho_{k,s_1} \tensor
  \dotsb \tensor \rho_{k,s_t}\}_{s_1,\dotsc,s_t \in \mathcal{S}}$ are
  mutually orthogonal. By putting this all together we conclude the proof. 
\end{proof}

\begin{cor}
  \label{cor:keysize}
  For a QSR scheme to be $(t,\epsilon)$-randomizing or
  $(t,\epsilon)$-in\-dis\-tin\-guish\-able, it is necessary that
  \begin{equation}
    \label{eq:keysize.cor}
    \Hh \left( \mathcal{K} \right) \geq (1 - 8 \epsilon) t \log d - 2,
  \end{equation}
  where $d$ is the dimension of the message Hilbert space $\hilbert_S$
  and $\Hh \left( \mathcal{K} \right)$ is the entropy of the
  decryption key.
\end{cor}

\begin{proof}
  \defref{def:model.classical.security} says that for a scheme to be
  $(t,\epsilon)$-secure we need \[\trnorm{\rho^{S^tC^t} - \rho^{S^t}
    \tensor \rho^{C^t}} \leq \epsilon\] for all probability
  distributions $P_{\mathcal{S}^t}$. So for the uniform distribution we
  get from \thmref{thm:keysize} that for a scheme to be
  $(t,\epsilon)$-secure we need \[ \Hh \left( \mathcal{K} \right) \geq
  (1 - 4 \epsilon) t \log |\mathcal{S}| - 2. \]

  By \lemref{lem:model.classical.security} we then have the
  condition \[ \Hh \left( \mathcal{K} \right) \geq (1 - 8 \epsilon) t
  \log |\mathcal{S}| - 2 \] for the scheme to be
  $(t,\epsilon)$-randomizing for the classical messages
  $\mathcal{S}$. And as classical messages are a subset of quantum
  messages -- namely an orthonormal basis of the message Hilbert space
  -- this bound extends to the case of quantum messages on a Hilbert
  space of dimension $d_S = |\mathcal{S}|$.

  As $(t,\epsilon)$-randomization is a special case of
  $(t,\epsilon)$-indistinguishability, namely for $t_1 = t$, it is
  immediate that this lower bound also applies to
  $(t,\epsilon)$-in\-dis\-tin\-guish\-a\-bi\-li\-ty. 
\end{proof}

\begin{rem}
  \label{rem:keysize}
  Approximate quantum one-time pad schemes usually only consider the
  special case in which the cipher has the same dimension as the
  message~\cite{HLSW04,DN06}. A more general scenario in which an
  ancilla is appended to the message is however also possible. It was
  proven in \cite{NS07} that for perfect security such an extended
  scheme needs a key of the same size as in the restricted scenario,
  namely $2 \log d$. \corref{cor:keysize} for $t = 1$ shows the same
  for approximate security, namely roughly $\log d$ bits of key are
  necessary, just as when no ancilla is present.
\end{rem}

\section{Near-Optimal Scheme}
\label{sec:scheme}

To simplify the presentation of the QSR scheme, we first define it for
classical messages in \secref{sec:scheme.classical}, show that it is
invertible and find a bound on $t$, the number of copies of the
encryption key which can be released, for it to be
$(t,\epsilon)$-randomizing for an exponentially small $\epsilon$. In
\secref{sec:scheme.quantum} we extend the scheme to encrypt any
quantum message of a given size, and show again that it is invertible
and randomizing. And finally in \secref{sec:scheme.key} we calculate
the size of the key necessary to encrypt a message of a given length,
and show that it is nearly asymptotically equal to the lower bound
found in \secref{sec:keysize}.

\subsection{Classical Messages}
\label{sec:scheme.classical}

Without loss of generality, let the message space be of
dimension $\dim \hilbert_S = 2^m$. The classical messages can then be
represented by strings of length $m$, $\mathcal{S} \coloneqq
\{0,1\}^m$. We now define a QSR scheme which uses encryption key
states of dimension $\dim \hilbert_K = 2^{m+n}$, where $n$ is a
security parameter, i.e., the scheme will be
$(t,\epsilon)$-randomizing for $\epsilon = 2^{-\Theta(n)}$.

We define the set of decryption keys to be the set of all $(m \times
n)$ binary matrices, \begin{equation} \label{eq:scheme.classical.deckey}
  \mathcal{K} \coloneqq \{0,1\}^{m \times n}.\end{equation} This set
has size $|\mathcal{K}| = 2^{mn}$ and each key is chosen with uniform
probability.

For every decryption key $A \in \mathcal{K}$ the corresponding
encryption key is defined as
\begin{equation} \label{eq:scheme.classical.enckey} \rho_A \coloneqq
  \frac{1}{2^n} \sum_{x \in \{0,1\}^n} \proj{Ax,x},\end{equation}
where $Ax$ is the multiplication of the matrix $A$ with the vector
$x$.

The encryption operator $\mathcal{E} : \mathcal{B}(\hilbert_K \tensor
\hilbert_S) \rightarrow \mathcal{B}(\hilbert_C)$ consists in applying
the unitary \[U \coloneqq \sum_{\substack{x \in \{0,1\}^n \\ s,y \in
    \{0,1\}^m}} \ket{y \oplus s,x}\bra{y,x}^K \proj{s}^S\] and tracing
out the message system $S$, i.e., \[ \rho_{A,s} \coloneqq
\trace[S]{U\left(\rho_k^K \tensor \proj{s}^S \right) \hconj{U}}.\]
This results in the cipher for the message $s$ being
\begin{equation} \label{eq:scheme.classical.encop}\rho_{A,s} = \frac{1}{2^n} \sum_{x \in \{0,1\}^n} \proj{Ax
  \oplus s,x}.\end{equation} These states are mutually orthogonal for different
messages $s$ so by \remref{rem:model.classical.dec} this scheme is
invertible.

We now show that this scheme is $(t,\epsilon)$-randomizing for
$\epsilon = 2^{- \delta n + 1}$ and $t = (1 - \delta) n$, $0 < \delta <
1$.

\begin{thm}
  \label{thm:scheme.classical}
  For the QSR scheme defined above in
  \eqnsref{eq:scheme.classical.deckey},
  \eqref{eq:scheme.classical.enckey} and
  \eqref{eq:scheme.classical.encop} there exists a density operator
  $\tau \in \mathcal{B}(\hilbert_C^{\tensor t})$ such that for all
  $t$-tuples of messages $s = (s_1,\dotsc,s_t) \in \mathcal{S}^{\times
    t}$, if $t = (1 - \delta) n$, $0 < \delta < 1$, then \[
  \trnorm{\gamma_s - \tau} \leq 2^{- \delta n + 1},\] where $\gamma_s$
  is the encryption of $s$ with this scheme averaged over all possible
  keys,
  i.e., \begin{equation} \label{eq:scheme.classical.thm.gammas}
    \gamma_s = \sum_{A \in \mathcal{K}} P_{\mathcal{K}}(A)
    \rho_{A,s_1} \tensor \dotsb \tensor \rho_{A,s_t}.\end{equation}
\end{thm}

\begin{proof}
  The $\tau$ in question is the fully mixed state $\tau =
  \frac{1}{2^{t(m+n)}} \id$. By placing the values of the ciphers from
  \eqnref{eq:scheme.classical.encop} in
  \eqnref{eq:scheme.classical.thm.gammas} we get \[ \gamma_s =
  \frac{1}{2^{mn}2^{tn}} \sum_{\substack{A \in \{0,1\}^{m \times n} \\
      x_1,\dotsc,x_t \in \{0,1\}^n}} \proj{\dotsc, Ax_i \oplus
    s_i,x_i,\dotsc}.\] A unitary performing bit
  flips can take $\gamma_s$ to $\gamma_r$ for any $s,r \in
  \mathcal{S}^t$, so \[ \trnorm{\gamma_s - \frac{1}{2^{t(m+n)}} \id}
= \trnorm{\gamma_r - \frac{1}{2^{t(m+n)}} \id},\] and it is sufficient
to evaluate
  \begin{equation}
    \label{eq:scheme.classical.thm.eigs}
    \trnorm{\gamma_0 -
      \frac{1}{2^{t(m+n)}} \id} = \sum_{e \in \text{EVec}(\gamma_0)} \left| w_e -
      \frac{1}{2^{t(n+m)}} \right|,
  \end{equation}
  where $e$ are the eigenvectors of $\gamma_0$ and $w_e$ the
  corresponding eigenvalues.

  So we need to calculate the eigenvalues of
  \begin{equation}
    \label{eq:scheme.classical.thm.gamma0}
    \gamma_0 = \frac{1}{2^{mn}2^{tn}} \sum_{\substack{A \in \{0,1\}^{m \times n} \\ x_1,\dotsc,x_t \in \{0,1\}^n}} \proj{Ax_1,x_1,\dotsc,Ax_t,x_t}.
  \end{equation}
  Let us fix $x_1,\dotsc,x_t$. It is immediate from the linearity of
  $Ax$ that if exactly $d$ of the vectors $\{x_i\}_{i=1}^t$ are
  linearly independent, then \[\sum_{A \in \{0,1\}^{m \times n}}
  \proj{Ax_1,x_1,\dotsc,Ax_t,x_t}\] uniformly spans a space of
  dimension $2^{dm}$, and for different values of $x_1,\dotsc,x_t$
  these subspaces are all mutually orthogonal. Let $D_t$ be the random
  variable representing the number of independent vectors amongst $t$
  binary vectors of length $n$, when chosen uniformly at random, and
  let $P_{D_t}(d) = \Pr[D_t = d]$ be the probability that exactly $d$
  of these vectors are linearly independent. The matrix given in
  \eqnref{eq:scheme.classical.thm.gamma0} then has exactly
  $2^{tn}P_{D_t}(d)2^{dm}$ eigenvectors with eigenvalue
  $\frac{1}{2^{dm}2^{tn}}$, for $0 \leq d \leq t$. The remaining
  eigenvectors have eigenvalue $0$.

  So \eqnref{eq:scheme.classical.thm.eigs} becomes 
  \begin{align*}
    \sum_{e \in \text{EVec}(\rho^E_0)} \left| w_e - \frac{1}{2^{t(m+n)}}\right| & =
    2 \sum_{d = 0}^t 2^{tn}P_{D_t}(d)2^{dm} \left( \frac{1}{2^{dm}2^{tn}} - \frac{1}{2^{t(m+n)}}\right) \\
    &  =  2 \sum_{d = 0}^t P_{D_t}(d) \left(1 - 2^{-(t-d)m}\right) \\
    &  \leq 2 \sum_{d = 0}^{t-1} P_{D_t}(d) = 2 (1 -  P_{D_t}(t)) \\
    &  \leq 2^{t-n+1}.
  \end{align*}
  For $t = (1 - \delta) n$, $0 < \delta < 1$, we have for all $s \in
  \mathcal{S}^t$, $\trnorm{\gamma_s - \tau} \leq 2^{-\delta n+1}$. 
\end{proof}

\begin{cor}
  \label{cor:scheme.classical}
  An asymmetric-key cryptosystem using this QSR scheme is
  $(t,\epsilon)$-indistinguishable (\defref{def:model.cryptosystem})
  for $\epsilon = 2^{- \delta n + 1}$ and $t = (1 - \delta) n$, $0 <
  \delta < 1$.
\end{cor}

\begin{proof}
  As noted in \secref{sec:model.cryptosystem} this scheme is such
  that the encryption keys are identical to the ciphers of the message
  $0$, $\rho_{k,0} = \rho_k = \frac{1}{2^n} \sum_{x \in \{0,1\}^n}
  \proj{Ax,x}$. So if $t_1$ copies of the encryption key were used to
  encrypt the messages $s = (s_1,\dotsc,s_{t_1})$ and the adversary
  holds these ciphers and the $t_2 = t - t_1$ extra copies of the
  encryption key, \[\rho^E_s = \sum_{k \in \mathcal{K}}
  P_{\mathcal{K}}(k) \rho_{k,s_1} \tensor \dotsb \tensor
  \rho_k,s_{t_1} \tensor \rho_k^{\tensor t_2}.\] Then $\rho_s^E =
  \gamma_r$ for $r = (s_1,\dotsc,s_{t_1},0,\dotsc,0)$ and by
  \thmref{thm:scheme.classical}, $\trnorm{\gamma_r - \tau} \leq
  \epsilon$.
\end{proof}

\subsection{Quantum Messages}
\label{sec:scheme.quantum}

We will now extend the encryption scheme given above to encrypt any
quantum state, not only classical ones. To do this we will show how to
combine a QSR scheme with quantum keys which is
$(t,\epsilon_1)$-randomizing for classical messages (like the one from
\secref{sec:scheme.classical}) with a QSR scheme with classical keys
which is $(1,\epsilon_2)$-randomizing for quantum states (which is the
case of any standard QSR scheme, e.g.,
\cite{BR03,AMTW00,HLSW04,AS04,DN06,NS07}) to produce a QSR scheme
which is $(t,\epsilon_1 + t \epsilon_2)$-randomizing. The general idea
is to choose a classical key for the second scheme at random, encrypt
the quantum message with this scheme, then encrypt the classical key
with the quantum encryption key of the first scheme, and send both
ciphers.

\begin{thm}
  \label{thm:scheme.quantum}
  Let a QSR scheme with quantum keys be given by the tuple
  $\mathbb{T}_1 = \left(P_\mathcal{K}, \{\rho_k\}_{k \in \mathcal{K}},
    \mathcal{E} \right)$, where $\mathcal{E} : \mathcal{B}(\hilbert_K
  \tensor \hilbert_S) \rightarrow \mathcal{B}(\hilbert_C)$, and let a
  QSR scheme with classical keys be given by the tuple $\mathbb{T}_2 =
  \left(P_\mathcal{S}, \{\mathcal{F}_s\}_{s \in \mathcal{S}}\right)$,
  where $\mathcal{F}_s : \mathcal{B}(\hilbert_R) \rightarrow
  \mathcal{B}(\hilbert_D)$. We combine the two to produce the QSR
  scheme with quantum encryption keys given by $\mathbb{T}_3 =
  \left(P_\mathcal{K}, \{\rho_k\}_{k \in \mathcal{K}}, \mathcal{G}
  \right)$, where $\mathcal{G} : \mathcal{B}(\hilbert_K \tensor
  \hilbert_R) \rightarrow \mathcal{B}(\hilbert_C \tensor \hilbert_D)$
  is defined by
  \begin{equation}
    \label{eq:scheme.quantum.encop}
    \mathcal{G}(\rho_k \tensor \sigma) \coloneqq \sum_{s \in \mathcal{S}}P_{\mathcal{S}}(s) \mathcal{E}\left(\rho_k \tensor \proj{s} \right) \tensor \mathcal{F}_s(\sigma).
  \end{equation}
  If $\mathbb{T}_1$ forms a quantum asymmetric-key cryptosystem which
  is invertible and $(t,\epsilon_1)$-indistinguishable (respectively
  randomizing) for the basis states of $\hilbert_S$ and
  $\mathbb{T}_2$ is an invertible and $(1,\epsilon_2)$-randomizing QSR
  scheme for any state on $\hilbert_R$, then $\mathbb{T}_3$ forms an
  invertible and $(t,\epsilon_1 + t \epsilon_2)$-indistinguishable
  (respectively randomizing) cryptosystem for all density operator
  messages on $\hilbert_R$.
\end{thm}

\begin{proof}
  The invertibility of the scheme formed with $\mathbb{T}_3$ is
  immediate. To prove the indistinguishability we need to show that
  for all $t_1 \in \{0,1,\dotsc,t\}$, $t_2 \coloneqq t - t_1$, there
  exists a density operator $\tau \in
  \mathcal{B}\left(\hilbert_C^{\tensor t_1} \tensor
    \hilbert_K^{\tensor t_2} \tensor \hilbert_D^{\tensor t_1} \right)$
  such that for all $t_1$-tuples of message density operators $\omega
  = (\sigma_1,\dotsc,\sigma_{t_1}) \in \mathcal{B}(\hilbert_R)^{\times
    t_1}$, $\trnorm{\rho^E_\omega - \tau} \leq \epsilon$, where
  $\rho^E_\omega = \sum_{k \in \mathcal{K}} P_{\mathcal{K}}(k)
  \mathcal{G}(\rho_k \tensor \sigma_1) \tensor \dotsb \tensor
  \mathcal{G}(\rho_k \tensor \sigma_{t_1}) \tensor \rho_k^{t_2}$.

  Let us write $\gamma_s \coloneqq \sum_{k \in \mathcal{K}}
  P_{\mathcal{K}}(k) \rho_{k,s_1} \tensor \dotsb \tensor
  \rho_{k,s_{t_1}} \tensor \rho_k^{\tensor t_2}$, where $s = (s_1,
  \dotsc, s_{t_1})$ and $\rho_{k,s_i} = \mathcal{E}(\rho_k \tensor
  \proj{s_i})$, and $\mu_{\sigma} \coloneqq \sum_{s \in \mathcal{S}}
  P_{\mathcal{S}}(s) \mathcal{F}_s(\sigma)$. And let $\tau_1$ and
  $\tau_2$ be the two states such that $\trnorm{\gamma_s - \tau_1}
  \leq \epsilon_1$ and $\trnorm{\mu_\sigma - \tau_2} \leq \epsilon_2$
  for all $s$ and $\sigma$ respectively. We define $\delta_s \coloneqq
  \gamma_s - \tau_1$ and $\tau \coloneqq \tau_1 \tensor
  \tau_2^{\tensor t_1}$. Then by the triangle inequality and changing
  the order of the registers
  \begin{align*}
    \trnorm{\rho^E_\omega - \tau} & \leq \trnorm{\sum_{s \in
        \mathcal{S}^{\times t_1}} P_{\mathcal{S}^t}(s) \delta_{s }
      \tensor \mathcal{F}_{s_1}(\sigma_1) \tensor \dotsb \tensor
      \mathcal{F}_{s_{t_1}}(\sigma_{t_1})} \\ & \quad + \trnorm{\tau_1 \tensor \mu_{\sigma_1} \tensor \dotsb \tensor \mu_{\sigma_{t_1}} - \tau} \\
    & \leq \sum_{s \in \mathcal{S}^{\times t_1}} P_{\mathcal{S}^t}(s)
    \trnorm{\delta_{s}} + \sum_{i = 1}^{t_1} \trnorm{\mu_{\sigma_i} -
      \tau_2} \\ & \leq \epsilon_1 + t_1 \epsilon_2.
  \end{align*}

  As $(t,\epsilon)$-randomization is a special case of
  $(t,\epsilon)$-indistinguishability, namely for $t_1 = t$, it is
  immediate from \thmref{thm:scheme.quantum} that $\mathbb{T}_3$ is
  also $(t,\epsilon_1 + t \epsilon_2)$-randomizing. 
\end{proof}

\subsection{Key Size}
\label{sec:scheme.key}

To construct the QSR scheme for quantum messages as described in
\secref{sec:scheme.quantum} we combine the scheme for classical
messages from \secref{sec:scheme.classical} and the approximate
one-time pad scheme of Dickinson and Nayak~\cite{DN06}.

The scheme from \secref{sec:scheme.classical} is
$(t,\epsilon_1)$-randomizing for $t = (1 - \delta)n$ and $ \epsilon_1
= 2^{-\delta n + 1}$, and uses a key with entropy $\Hh \left(
  \mathcal{K} \right) = nm = (t + \log \frac{1}{\epsilon_1} +
1)m$. The scheme of Dickinson and Nayak~\cite{DN06} is
$(1,\epsilon_2)$-randomizing and uses a key with entropy $m = \log d +
\log \frac{1}{\epsilon_2} + 4$ to encrypt a quantum state of dimension
$d$. So by combining these our final scheme is
$(t,\epsilon_1 + t \epsilon_2)$-randomizing and uses a key with entropy
\[ \Hh \left( \mathcal{K} \right) = (t + \log \frac{1}{\epsilon_1} +
1)(\log d + \log \frac{1}{\epsilon_2} + 4)\] to encrypt $t$ states of
dimension $d$. By choosing $\epsilon_1$ and $\epsilon_2$ to be
polynomial in $\frac{1}{t}$ and $\frac{1}{\log d}$ respectively, the
key has size $\Hh \left( \mathcal{K} \right) = t \log d + o(t \log
d)$, which nearly reaches the asymptotic optimality found in
\eqnref{eq:keysize.cor}, namely $ \Hh \left( \mathcal{K} \right) \geq
(1 - 8 \epsilon) t \log d - 2$. Exponential security can be achieved at
the cost of a slightly reduced asymptotic efficiency. For $\epsilon_1
= 2^{-\delta_1 t}$ and $\epsilon_2 = d^{- \delta_2}$ for some small
$\delta_1,\delta_2 > 0$, the key has size $\Hh \left( \mathcal{K}
\right) = (1+\delta_1)(1+\delta_2) t \log d + o(t \log d)$.

\section{Consequence for Quantum Keys}
\label{sec:consequence}

The scheme presented in \secref{sec:scheme} uses the encryption
keys \begin{equation} \label{eq:consequence.enckey} \rho_A =
  \frac{1}{2^n} \sum_{x \in \{0,1\}^n} \proj{Ax,x},\end{equation} for
some $(m \times n)$-matrix decryption key $A$. Although these keys are
written as quantum states using the bra-ket notation to fit in the
framework for QSR schemes with quantum keys developed in the previous
sections, the states from \eqnref{eq:consequence.enckey} are all diagonal in
the computational basis. So they are classical and could have been
represented by a classical random variable $\mathcal{X}_A$ which takes
the value $(Ax,x)$ with probability $2^{-n}$. 

This scheme meets the optimality bound on the key size from
\secref{sec:keysize}. This bound tells us that for a given set of
decryption keys $\mathcal{K}$, no matter how the encryption keys
$\{\rho_k\}_{k \in \mathcal{K}}$ are constructed, the number of copies
of the encryption keys which can be created, $t$, and the dimension of
the messages which can be encrypted, $d$, have to be such that $t \log
d \lesssim \Hh \left(\mathcal{K}\right)$ for the scheme to be
information-theoretically secure. From the construction of the scheme
in \secref{sec:scheme} we know that this bound is met by a scheme
using classical keys. Hence no scheme using quantum keys can perform
better. So using quantum keys in a quantum state randomization scheme
has no advantage with respect to the message size and number of usages
of the same key over classical keys.

This result applies to both the symmetric-key and asymmetric-key
models as the optimality was shown with respect to both
$(t,\epsilon)$-randomization (\defref{def:model.quantum.security}) and
$(t,\epsilon)$-indistinguishability (\defref{def:model.cryptosystem}),
the security definitions for the sym\-met\-ric-\-key and asymmetric-key
models respectively.

Quantum keys may however have other advantages over classical
keys. For example, the scheme proposed in \secref{sec:scheme} is not
optimal in the dimension of the encryption keys $\rho_k$. If the
dimension of these keys can be reduced and quantum memory becomes the
norm, they could be less resource consuming than classical keys. So
encryption schemes using quantum keys cannot yet be dismissed.

\section*{Acknowledgements}
\label{sec:acknowledgements}

The authors thank Renato Renner for helpful suggestions, in particular for the
proof of \thmref{thm:keysize}.

This work is partially supported by the Ministry of Education,
Science, Sports and Culture, Grant-in-Aid for Young Scientists (B)
No.17700007, 2005 and for Scientific Research (B) No. 18300002, 2006.

\bibliographystyle{eprintunsrt} 
\bibliography{quantum,Akinori}





  
\end{document}